\documentclass[conference]{IEEEtran}

\usepackage[top  =1in, left =0.75 in, right =0.75 in, bottom =0.75 in]{geometry}

\usepackage{amsmath, amssymb,bm}
\usepackage{enumerate,mathtools}
\usepackage{amsthm}
\usepackage{cite}
\usepackage{microtype}
\usepackage{booktabs}

\usepackage{tikz}
\usetikzlibrary{shapes.geometric, shapes.arrows, arrows, positioning}
\tikzstyle{box}=[draw, fill=blue!20, scale= 0.8, minimum size=2em]
\tikzstyle{circ}=[draw, circle, fill=red!20, scale= 0.8, minimum size=2.5em]
\tikzstyle{circ_blue}=[draw, circle, fill=blue!20, scale= 0.8, minimum size=2em]
\tikzstyle{circ_white}=[draw, circle, scale= 0.8, minimum size=2.5em]
\tikzstyle{box_red}=[draw, fill=red!20, minimum size=2em, scale= 0.8]

\usepackage{pgfplots}
\usepgfplotslibrary{fillbetween}

\theoremstyle{definition}

\newtheorem{theorem}{Theorem}

\theoremstyle{definition}
\newtheorem{definition}{Definition}

\newtheorem{assumption}{Assumption}

\newtheorem{example}{Example}

\newcommand{\bA}{\bm{A}}

\newcommand{\bK}{\bm{K}}

\newcommand{\bS}{\bm{S}}

\newcommand{\bU}{\bm{U}}
\newcommand{\bV}{\bm{V}}
\newcommand{\bW}{\bm{W}}
\newcommand{\bX}{\bm{X}}
\newcommand{\bY}{\bm{Y}}
\newcommand{\bZ}{\bm{Z}}

\newcommand{\bs}{\bm{s}}

\newcommand{\bu}{\bm{u}}
\newcommand{\bv}{\bm{v}}

\newcommand{\bx}{\bm{x}}
\newcommand{\by}{\bm{y}}
\newcommand{\bz}{\bm{z}}

\newcommand{\cF}{\mathcal{F}}

\global\long\def\dd{\mathrm{d}}

\newcommand{\ex}[1]{\ensuremath{\mathbb{E}\left[ #1\right]}}

\DeclareMathOperator{\cov}{\mathsf{Cov}}
\DeclareMathOperator{\mmse}{\sf mmse}

\DeclareMathOperator{\gtr}{tr}

\newcommand{\reals}{\mathbb{R}}

\newcommand{\eps}{\epsilon}
\newcommand{\normal}{\mathcal{N}}
\newcommand{\gmid}{\! \mid \!}

\let\originalleft\left
\let\originalright\right
\renewcommand{\left}{\mathopen{}\mathclose\bgroup\originalleft}
\renewcommand{\right}{\aftergroup\egroup\originalright}

\newcommand{\ST}{C} 	
\newcommand{\FR}{R} 	
\newcommand{\IR}{J} 	

\newcommand{\IM}{G} 	

\newcommand{\noise}{\bW} 
\newcommand{\haar}{\bU} 

  \newif\iflongpaper
    \longpaperfalse

  \newif\ifshowfigs

   \showfigstrue

\IEEEoverridecommandlockouts

\title{Additivity of Information in Multilayer Networks via Additive Gaussian Noise Transforms}

\author{Galen Reeves\\
Department of ECE and Department of Statistical Science\\ Duke University
\thanks{The work of G.\ Reeves was supported in part by funding from the Laboratory for Analytic Sciences (LAS).  Any opinions, findings, conclusions, and recommendations expressed in this material are those of the author and do not necessarily reflect the views of the sponsors. }
}
\begin{document}

 \maketitle

\begin{abstract}
Multilayer (or deep) networks are powerful probabilistic models based on multiple stages of a linear transform followed by a non-linear (possibly random) function. In general, the linear transforms are defined by matrices and the non-linear functions are defined by information channels. These models have gained great popularity due to their ability to characterize complex probabilistic relationships arising in a wide variety of inference problems. The contribution of this paper is a new method for analyzing the fundamental limits of statistical inference in settings where the model is known. The validity of our method can be established in a number of settings and is conjectured to hold more generally. A key assumption made throughout is that the matrices are drawn randomly from orthogonally invariant distributions. 

Our method yields explicit formulas for 1) the mutual information; 2) the minimum mean-squared error (MMSE); 3) the existence and locations of certain phase-transitions with respect to the problem parameters; and 4) the stationary points for the state evolution of approximate message passing algorithms. When applied to the special case of models with multivariate Gaussian channels our method is rigorous and has close connections to free probability theory for random matrices. When applied to the general case of non-Gaussian channels, our method provides a simple alternative to the replica method from statistical physics.  A key observation is that the combined effects of the individual components in the model (namely the matrices and the channels) are additive when viewed in a certain transform domain. 

\end{abstract}

\section{Introduction} 
Probabilistic models  involving high-dimensional linear transforms arise in a wide variety of applications throughout science and engineering. A canonical building block for these models can be understood in terms of the generalized linear model (GLM), which can be described as follows: 
\begin{align}
\bX \sim P_1(\bx) \, \qquad \bZ = \bA \bX, \qquad \bY \mid \bZ \sim P_2(\by \mid \bz) .
\end{align}
In this model, $P_1(\bx)$ is the prior distribution on the vector of unknown variables, $\bA$ is a known $M \times N$ matrix, and $P_2(\by \mid \bz)$ is an information channel (or likelihood) describing the conditional distribution of the vector of observations. This characterization of the GLM is quite broad. For applications in communication systems and compressed sensing, the channel $P(\by \gmid \bz)$ can be used to models additive noise or quantization error due to bitrate constraints. For applications in machine learning the channel can be used to model deterministic functions, such as a thresholding or pooling operation. 

\begin{figure}[t]
\center
\scalebox{.9}{
\begin{tikzpicture}[>= latex']
	
	\node (X1) at (0,0)  [circ_white, scale = 0.8]{$\bX_1$};
	\node (G2) [box, above right = .3 and .75 cm of X1, scale = 0.8]{$\text{GLM}_2$};
	\node (G3) [box, below right = .3  and .75 cm of X1, scale = 0.8]{$\text{GLM}_3$};

	\node (X2) [circ_white, right =  .7 cm of G2, scale = 0.8]{$\bX_{2}$};
	\node (X3) [circ_white, right =  .7 cm of G3, scale = 0.8]{$\bX_{3}$};
	
	\node (G4) [box, above right = .1 and .75 cm of X2, scale = 0.8]{$\text{GLM}_4$};
	\node (G5) [box, below right  = -.15 and .75 cm of X2, scale = 0.8]{$\text{GLM}_5$};
	\node (G6) [box, above right = -.15 and .75 cm of X3, scale = 0.8]{$\text{GLM}_6$};
	\node (G7) [box, below right  = .1 and .75 cm of X3, scale = 0.8]{$\text{GLM}_7$};

	\node (X4) [circ_white, right =  .7 cm of G4, scale = 0.8]{$\bX_{4}$};
	\node (X5) [circ_white, right =  .7 cm of G5, scale = 0.8]{$\bX_{5}$};
	\node (X6) [circ_white, right =  .7 cm of G6, scale = 0.8]{$\bX_{6}$};
	\node (Y7) [circ, right =  .7 cm of G7, scale = 0.8]{$\bY_{7}$};
	
	\node (G8) [box, right =  .75 cm of X4, scale = 0.8]{$\text{GLM}_8$};
	\node (G9) [box, right  = .75 cm of X6, scale = 0.8]{$\text{GLM}_9$};
	
	\node (Y8) [circ, right =  .7 cm of G8, scale = 0.8]{$\bY_{8}$};
	\node (Y9) [circ, right =  .7 cm of G9, scale = 0.8]{$\bY_{9}$};
	
	\foreach \i in {2,...,6}
	\draw [semithick,->] (G\i) -- (X\i);
	
	\foreach \i in {7,...,9}
	\draw [semithick,->] (G\i) -- (Y\i);
	
	\draw [semithick,->] (X1) -- (G2);
	\draw [semithick,->] (X1) -- (G3);
	\draw [semithick,->] (X2) -- (G4);
	\draw [semithick,->] (X2) -- (G5);
	\draw [semithick,->] (X3) -- (G6);
	\draw [semithick,->] (X3) -- (G7);
	\draw [semithick,->] (X4) -- (G8);
	\draw [semithick,->] (X6) -- (G9);
\end{tikzpicture}
}
\caption{\label{fig:TGLM} Example of a tree network of GLMs. The nodes indexed by $\{1,\dots,6\}$ correspond to unknown random  vectors and the nodes indexed by $\{7, 8, 9\}$ correspond to observations. Each GLM consists of an $M_\ell \times N_\ell$ matrix $\bA_\ell$ followed by an information channel $P_\ell(\cdot \mid \cdot)$. }
\vspace{-.2in}
\end{figure}
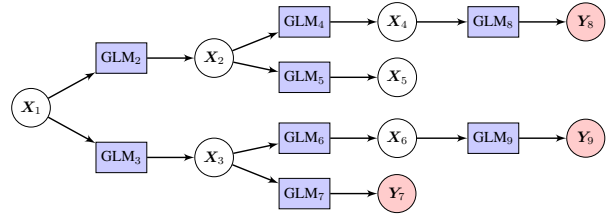

Multilayer (or deep) networks, such as the one illustrated in Figure~\ref{fig:TGLM}, can be viewed as the composition of multiple GLMs. In these networks, the channels are often highly decomposable in the sense that each output depends only on a small number of inputs.  By combining multiple stages of linear transforms with decomposable but non-linear functions, these models can capture complex probabilistic relationships. 

The dependencies induced by the linear transforms can give rise to fascinating phenomena in high-dimensional settings. For example,  the typical behavior of the posterior distribution can exhibit phase transitions in which a small change in the parameters of the model (such as the number of observations) can lead to massive differences in statistical  performance metrics, such as the probability of error in classification or the minimum mean squared error (MMSE) in estimation. In many cases, these phase transitions correspond to the boundaries between \textit{easy} problem regimes,  in which  efficient methods for approximate inference are essentially optimal, and \textit{hard} problem regimes,  in which there exists a large performance gap between optimal inference and all known efficient methods. 

One of the central challenges facing researchers is to understand the global behavior of these networks in terms of their individual components. In this direction, there has been significant contributions from a number of different disciplines. In the context of Gaussian networks, results from free probability theory and random matrix theory \cite{voiculescu:1991} have been used to study the fundamental limits of wireless communication systems~\cite{tulino:2004}.

\newgeometry{margin = .75in}

Non-Gaussian networks have been addressed using the heuristic replica method from statistical physics~\cite{mezard:2009}.  This approach yields precise formulas that are conjectured to be exact in the limit of large problem dimensions. The replica method has been applied to applications in wireless communications~\cite{tanaka:2002, guo:2005}, compressed sensing \cite{guo:2009, reeves:2012, tulino:2013}, and multilayer generalized linear models~\cite{manoel:2017a}. The main limitation, however,  is that the validity of formulas relies on certain key assumptions, most notably replica symmetry, which are unproven in general.  Recently, there has been a great deal of progress in providing rigorous results for certain non-Gaussian networks \cite{korada:2010,huleihel:2017, reeves:2016a, barbier:2016, barbier:2017c}. 
 
The global behavior of these networks can also be understood by studying the performance of specific algorithms such as approximate message passing (AMP)\cite{donoho:2009a,bayati:2011} and its generalizations \cite{rangan:2011,cakmak:2014a,  rangan:2016a, schniter:2016b, manoel:2017a,  fletcher:2017a}. In some cases, the behavior of these algorithms can be characterized precisely via a state evolution formalism,  which leads to single-letter characterizations of the behavior in the large system limit.  

\subsection{Contributions} 

The contribution of this paper is a new method for analyzing the statistical properties of multilayer networks. Our contributions include:
\begin{itemize}
\item Rigorous  formulas for mutual information and MMSE  in Gaussian networks with orthogonally invariant matrix distributions. This characterization provides a new estimation-theoretic perspective  on ideas from random matrix theory and free probability theory. 

\item Postulated formulas for mutual information and MMSE in non-Gaussian networks with orthogonally invariant matrix distributions. For a number of special cases, these formulas recover results that have been obtained previously using the replica method from statistical physics (see Table~\ref{tab:overview}). Two such examples are the linear model with orthogonally invariant distributions studied by Tulino et al.~\cite{tulino:2013} and serial GLM networks with IID Gaussian matrices studied by Manoel et al.~\cite{manoel:2017a}. More generally, this work is the first to provide explicit formulas for the case of arbitrary channels and arbitrary spectral distributions on the matrices. 

\item Definition of a potential function whose stationary points characterize the behavior of recent approximate message passing algorithms proposed by Schniter et al.~\cite{schniter:2016b}, Manoel et al.~\cite{manoel:2017a},  and Fletcher and Rangan~\cite{fletcher:2017a}. This result allows one to make explicit conjectures about the optimality of these algorithms based on whether the stationary point achieved by the algorithm is the unique global minimizer. 

\end{itemize}

Our method can be viewed as a simple alternative to the replica method in the sense that it provides a specific recipe for computing precise formulas. Similar to the replica method, the necessary and sufficient conditions under which these formulas  hold are not yet fully understood. However, an important distinction is that the main technical requirements for our approach depend only on the concentration of the information density and certain conditional central limit theorems, and thus bypass some of the notoriously difficult assumptions needed by the replica method. 

Due to space constraints, the paper outlines the main steps in the method.  The derivation follows in part from ideas used in the expectation consistent approximate inference framework of Opper and Winther~\cite{opper:2005} as well as Gaussian approximations for random projections studied by the author \cite{reeves:2017c}.

\subsection{Overview of main results}\label{sec:overview}  

Consider a GLM network such as the one illustrated in Figure~\ref{fig:TGLM}. Each GLM consists of an $M_\ell \times N_\ell$ matrix $\bA_\ell$ followed by an information channel $P_\ell(\cdot \gmid \cdot)$.  The key assumption made throughout is that these matrices are drawn randomly from orthogonally invariant distributions. Under this assumption, each matrix can be decomposed as
\begin{align}
\bA_\ell  =  \bU_\ell \bS_\ell \bV_\ell^T,
\end{align}
where $\bU_\ell, \bS_\ell, \bV_\ell$ are independent and $\bU_\ell$ and $\bV_\ell$ are uniformly distributed orthogonal matrices. 

\begin{table*}
\centering
\caption{\label{tab:overview}Overview of results. The algorithmic results correspond to mean-squared error obtained via the state evolution (SE) formalism. The fundamental limits correspond to explicit formulas for the mutual information and MMSE. } 
\begin{tabular}{lllll}
\toprule
& \multicolumn{2}{l}{IID Gaussian Matrices}  &\multicolumn{2}{l}{Orthogonally Invariant Matrices} \\
\cmidrule(r){2-5}
& Algorithmic SE &  Fundamental  Limits & Algorithmic SE &  Fundamental  Limits \\
\midrule
Linear Model  (Gaussian Case)  & AMP~\cite{donoho:2009a,bayati:2011}  & Rigorous\cite{voiculescu:1991, tse:1999,verdu:1999,tulino:2004} & S-AMP~\cite{cakmak:2014a}, VAMP~\cite{rangan:2016a} & Rigorous\cite{voiculescu:1991,tulino:2004} \\[.2em]
Linear Model  (IID Prior) &   AMP~\cite{donoho:2009a,bayati:2011} &  Postulated~\cite{guo:2005}, 
Rigorous~\cite{reeves:2016a} &S-AMP~\cite{cakmak:2014a},   VAMP~\cite{rangan:2016a} &Postulated~\cite{tulino:2013}   \\[.2em]
Generalized Linear Model &   GAMP~\cite{rangan:2011} &Postulated~\cite{manoel:2017a},  Rigorous~\cite{barbier:2017c}&   GVAMP~\cite{schniter:2016b} & \textbf{Postulated (This Paper)} \\[.2em]
Serial Network & ML-AMP~\cite{manoel:2017a} &Postulated~\cite{manoel:2017a} & ML-VAMP~\cite{fletcher:2017a} & \textbf{Postulated (This Paper)} \\[.2em]
Tree  Network (Gaussian Case)    & --  &\textbf{Rigorous (This Paper)}  & -- &\textbf{Rigorous  (This Paper)} \\
Tree  Network  & --  &\textbf{Postulated (This Paper)}  & -- &\textbf{Postulated  (This Paper)} \\
\bottomrule
\end{tabular}
\end{table*}

We focus on a sequence of problems, indexed by $N$,  in which the network is fixed, but the dimension of vectors and matrices increase to infinity. The asymptotic behavior of the model is described in terms of a potential function $\cF(\bu)$ where the  number of inputs  $\bu =\{u_\ell \}$ is equal to the number of unobserved vectors. The potential function is composed of the sum of terms, each of which depends only on an individual GLM in the network. Under certain assumptions, this potential has the following properties: 
\begin{itemize}
\item \textit{Mutual Information:} The mutual information between the unknown vectors and the observations corresponds to the minimum of the potential function:
\begin{align*}
\lim_{N \to \infty} \frac{1}{N} I( \underline{\bX} ;  \underline{\bY}  \mid  \underline{\bA}  ) & = \min_{\bu} \cF(\bu) .
\end{align*}
More generally, our method can also be used to characterize the mutual information corresponding to subsets of the variables and the observations. 
\item \textit{MMSE:} The minimum mean squared error (MMSE) of the unknown vectors corresponds to the minimizer of the potential function. In particular, if $\cF(\bu)$ as a unique global minimizer at $\bu^* = \{u_\ell^*\}$ then
\begin{align*}
\lim_{N \to \infty} \frac{1}{N} \mmse( \bX_\ell \mid  \underline{\bY}  ,  \underline{\bA}  ) & = u_\ell^*.
\end{align*}

\item \textit{Phase Transitions:} The system undergoes a phase transition with respect to a problem parameter (e.g., signal-to-noise ratio)  when the global minimum of the potential jumps from one local minimum to another. These phase transitions correspond to the locations where the minimum of the potential function is non-analytic with respect to perturbations of the paramter. 

\item \textit{Algorithmic Fixed-Points:} The stationary points of the potential function are the solutions to the equation
\begin{align*}
\nabla \cF(\bu) = \bm{0},
\end{align*}
where $\nabla$ is the gradient operator. The solutions of this equation correspond to  the fixed-points of the state evolution equations for AMP algorithms \cite{donoho:2009a,bayati:2011,rangan:2011,cakmak:2014a,   rangan:2016a, schniter:2016b, manoel:2017a,  fletcher:2017a} .  
\end{itemize}

\section{Additive Gaussian Noise Transforms}\label{sec:transforms}

\subsection{Mutual information and MMSE in Gaussian noise}

Let $\bX$ be an $N$-dimensional random vector. Given any number $s \in [0,\infty)$ we define  $\widetilde{\bX}(s)  = \sqrt{s} \bX + \noise$ to be an observation of $\bX$ under standard Gaussian noise $\noise \sim \normal(0, I)$ that is independent of everything else.  The mutual information function  $I_{\bX}(s)$ and MMSE function $M_{\bX}(s)$ are defined as
\begin{align*}
I_{\bX}(s) & = \frac{1}{N}  I\big(\bX ; \widetilde{\bX}(s)\big)\\
M_{\bX}(s) & = \frac{1}{N} \ex{ \gtr\left( \cov\big( \bX \mid \widetilde{\bX}(s)\big)\right)}.
\end{align*}
The MMSE function is finite and real analytic on $(0, \infty)$ \cite{guo:2011}. Furthermore, if the mutual information function is finite,  then its derivative is equal to one half the MMSE function: 
\begin{align}
\frac{\dd}{ \dd s}  I_{\bX}(s) &= \frac{1}{2}M_{\bX}(s). \label{eq:I_MMSE}
\end{align}
This identity is known as the I-MMSE relationship \cite{guo:2005a}.  
Combining the I-MMSE relationship  with the fact that the MMSE function is non-increasing, one finds that the mutual information function is concave. Noting that $I_{\bX}(0) = 0$ leads to an integral characterization of the  I-MMSE relationship: 
\begin{align}
I_{\bX}(s) & = \frac{1}{2} \int_0^s M_{\bX}(t) \, \dd t. \label{eq:I_MMSE_int}
\end{align}

The definitions given above can also be extended to the setting of a random pair $(\bX, \bY)$.  The conditional mutual information and MMSE functions associated with the conditional distribution of $\bX$ given $\bY$ are defined as
\begin{align*}
I_{\bX \mid \bY}(s) & = \frac{1}{N}  I\big(\bX ; \widetilde{\bX}(s) \mid \bY\big )\\
M_{\bX \mid \bY}(s) & = \frac{1}{N} \ex{ \gtr\left( \cov\big( \bX \mid \bY,  \widetilde{\bX}(s)\big)\right)}.
\end{align*}
These functions correspond to expectations over the joint distribution on $(\bX, \bY)$. They are linear in the marginal distribution of $\bY$ and satisfy the  I-MMSE relationship~\eqref{eq:I_MMSE}.  Furthermore, we define 
\begin{align*}
I_{\bX \triangle \bY}(s ) & =  \frac{1}{N}  I\big(\bX ;\bY,  \widetilde{\bX}(s) \big ),
\end{align*}
to be the mutual information between $\bX$ and the pair of observations $(\bY, \widetilde{\bX}(s))$. By the chain rule for mutual information,  
\begin{align}
I_{\bX \triangle \bY}(s ) & = I_{\bX \triangle \bY}(0)   +   I_{\bX \mid \bY}(s ), \label{eq:I_XY_decomp1}
\end{align}
and thus $I_{\bX \mid \bY}(s)$ and $I_{\bX \triangle \bY}(s)$ are equal up to a constant. Alternatively, using the  I-MMSE relationship \eqref{eq:I_MMSE_int} provides an integral characterization of mutual information in terms of the difference in the MMSE functions: 
\begin{align}
I_{\bX \triangle \bY}(s) & =  I_{\bX}(s) +   \frac{1}{2} \int_{s}^\infty \left(  M_{\bX}(s)  -  M_{\bX \mid \bY}(s ) \right) \,  \dd s. \label{eq:I_XY_decomp2}
\end{align}
This decomposition is illustrated graphically in Figure~\ref{fig:MI_functions}. Note that the first term depends only on the distribution of $\bX$ while the second term corresponds to the conditional mutual information $I(\bX; \bY \mid \widetilde{\bX}(s))$. Evaluating at $s=0$ recovers the characterization of mutual information given by Verd\'{u}~\cite[Theorem~3]{verdu:2010}.

\subsection{Legendre transforms} 

The previous section showed how the mutual information between arbitrary random vectors can be decomposed into estimation-theoretic quantities involving the difference between MMSE functions under additive Gaussian noise.  One of the key observations of this paper is that these decompositions can have useful properties when viewed in a transform domain. Specifically, we focus on the Legendre transform (or convex conjugate), which is defined according to
\begin{align*}
I^*_{\bX}(u) & = \sup_{s  \in [0, \infty) } \left( I_{\bX}(s)  - \frac{1}{2}  u s \right). 
\end{align*}
The factor of one half is used in this definition so that the parameter $u$ corresponds to the MMSE. The function $I^*_{\bX}(u)$ is a convex function over the domain $\{ u \in \reals  \, : \, I^*_{\bX}(u)  < \infty\}$. The fact that $I_{\bX}(s)$ is concave means that it can be recovered by applying the  Legendre transform a second time: 
\begin{align*}
I_{\bX}(s) & = \inf_{ u  } \left( I^*_{\bX}(u)  + \frac{1}{2}  s u \right),
\end{align*}
where the infimum is over the domain of $I^*_{\bX}(u)$. 

In the transform domain, the counterpart of the MMSE function is given by the inverse MMSE function $\IM_{\bX}(u)$, which is defined to be the functional inverse of $M_{\bX}(s)$. For any non-constant random vector it can be shown that the MMSE function is strictly decreasing on $[0, \infty)$. Therefore, the inverse is well-defined for all $u \in (0, M_{\bX}(0)]$ and is given by the unique solution to
\begin{align}
M_{\bX}( \IM_{\bX}(u)) = u.
\end{align} 
In words, $\IM_{\bX}(u)$ can be understood as the signal-to-noise ratio that is needed in order to attain a desired MMSE.  

The I-MMSE relationship in the transformed domain can now be stated as:
\begin{align}
\frac{\dd}{ \dd u}  I^*(u) &=  -  \frac{1}{2} \IM_{\bX}(u). 
\end{align}
Moreover, noting that $I_{\bX}^*(M_{\bX}(0)) = 0$ gives 
\begin{align}
 I_{\bX} ^*(u) &=  \frac{1}{2} \int_u^{M_{\bX}(0)}  \IM_{\bX}(v) \, \dd v.
\end{align}
From these expressions, it is easy to see  that an equivalent definition of the Legendre transform is provided by
\begin{align*}
I^*_{\bX}(u) & =    I_{\bX}(\IM(u) )  - \frac{1}{2} u  \,  \IM(u), \qquad 0 < u  \le  M_{\bX}(0).
\end{align*}
The behavior as $u$ converges to zero depends on whether $\bX$ has finite entropy:
\begin{align}
\lim_{u \to 0}   I^*_{\bX}(u)  & =
\begin{dcases}
\frac{1}{N} H(\bX),  & H(\bX) < \infty\\
 + \infty, & \text{otherwise} .
 \end{dcases}
\end{align}
   
\begin{figure}
\scalebox{.85}{
\begin{tikzpicture}
\begin{axis}
[thick,
  axis x line=bottom,
  axis y line=left,
    xtick={0},
     extra x ticks={2},
      extra x tick labels={$s$},
extra y ticks={0.333, 0.6, 1},
extra y tick labels={$$, $\scriptstyle M_{\bX \mid\bY}(0)$, $\scriptstyle M_{\bX}(0)$},
  ytick={0},
  xlabel style={below},
  ylabel style={rotate=-90},
  xmin=-0,
  xmax=5.5,
  ymin=-0,
  ymax=1.05]
  
  \addplot[name  path= m1, domain = .01:8, samples = 100, black, ultra thick] {1/(1+x)};

  \addplot[name  path= m2, domain = 0:8, samples = 150, smooth,  black , ultra thick, dash pattern={on 7pt off 2pt on 1pt off 3pt}] {.6/(1+  x ) };

        \legend{$M_{\bX}(s)$,$M_{\bX \mid \bY}(s)$}

\draw[thick, gray, dashed] (axis cs:2,.0) --(axis cs:2,.333);

\path[name path=axis] (axis cs:0,0) -- (axis cs:4,0);
 \path[name path=uline] (axis cs:0,.333) -- (axis cs:4,.333);

 \addplot [
        thick,
        color=red,
        fill=blue, 
        fill opacity=0.2
    ]
    fill between[of=m1 and axis, soft clip={domain=0:2}];

     \addplot [
        thick,
        color=red,
        fill=red, 
        fill opacity=0.2
    ]
    fill between[of=m1 and m2, soft clip={domain=2:6}];

\end{axis}

\end{tikzpicture}
}

\scalebox{.85}{
\begin{tikzpicture}
\begin{axis}
[thick,
  axis x line=bottom,
  axis y line=left,
    xtick={0},
     extra x ticks={2},
            extra x tick labels={$$},
extra y ticks={0.333, 0.6, 1},
extra y tick labels={$u$, $\scriptstyle M_{\bX \mid\bY}(0)$, $\scriptstyle M_{\bX}(0)$},
  ytick={0},
  xlabel style={below},
  ylabel style={rotate=-90},
  xmin=-0,
  xmax=5.5,
  ymin=-0,
  ymax=1.05]
  
  \addplot[name  path= m1, domain = .01:8, samples = 100, black, ultra thick] {1/(1+x)};

  \addplot[name  path= m2, domain = 0:8, samples = 150, smooth,  black , ultra thick, dash pattern={on 7pt off 2pt on 1pt off 3pt}] {.6/(1+  x ) };   

\draw[thick, gray,  dashed] (axis cs:0,.333) --(axis cs:2,.333);

\path[name path=axis] (axis cs:0,0) -- (axis cs:4,0);
 \path[name path=uline] (axis cs:0,.333) -- (axis cs:4,.333);

   \legend{$M_{\bX}(s)$,$M_{\bX \mid \bY}(s)$}

 \addplot [
        thick,
        color=red,
        fill=blue, 
        fill opacity=0.2
    ]
    fill between[of=m1 and uline, soft clip={domain=0:2}];

     \addplot [
        thick,
        color=red,
        fill=red, 
        fill opacity=0.2
    ]
    fill between[of=m2 and uline, soft clip={domain=.8:2}];

     \addplot [ thick,  fill=red, fill opacity=0.2] fill between[of=m1 and m2, soft clip={domain=2:4}];

\end{axis}
\end{tikzpicture}
}

\caption{\label{fig:MI_functions}Illustration of mutual information functions. The function $I_{\bX \triangle \bY}(s)$ is equal to one half of the area of the shaded regions shown in the top panel. The region on the left (blue) depends only on the prior distribution of $\bX$ while the region on the right (red) corresponds to the difference between MMSE functions. The function $I^*_{\bX \triangle \bY}(u)$ is equal to one half of the area of the shaded regions shown in the bottom panel. The top region (blue) depends only on the prior distribution of $\bX$ while the region on the bottom corresponds to the difference between inverse MMSE functions. 
}
\end{figure}
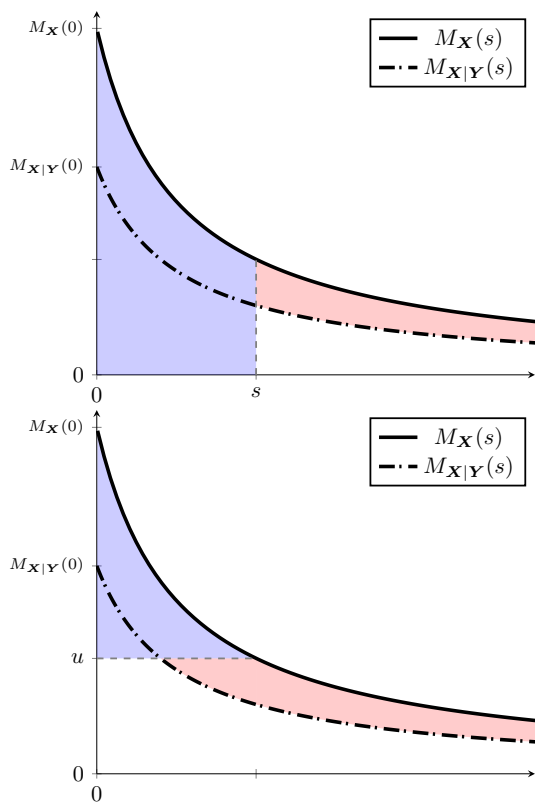

For a random pair $(\bX, \bY)$ the transformations associated with the conditional distribution of $\bX$ given $\bY$ are defined similarly. 
Following from \eqref{eq:I_XY_decomp1}, we see that
\begin{align}
I^*_{\bX \triangle \bY}(u) & = I_{\bX \triangle \bY}(0)  +  I^*_{\bX \mid \bY}(u), 
\end{align}
and thus $I^*_{\bX \triangle \bY}(u) $ and $I^*_{\bX \mid \bY}(u)$ are equal up to a constant. Alternatively, using the I-MMSE relationship the counterpart of the decomposition in \eqref{eq:I_XY_decomp2} is given by
\begin{align}
I^*_{\bX \triangle \bY}(u)
& =  I_{\bX}^*(u) + \frac{1}{2} \int_0^u  \left( \IM_{\bX }(v)  - \IM_{\bX \mid \bY }(v) \right) \, \dd v. \label{eq:Istar_XY_decomp2}
\end{align}
This decomposition is illustrated graphically in Figure~\ref{fig:MI_functions}. Note that the first term depends only on the distribution of $\bX$. The second term is non-negative and plays an important role in our analysis.

\subsection{Multivariate Legendre transforms} 

The definitions given in the previous subsections can also be extended to a collection of jointly random vectors $\underline{\bX} = \{\bX_1, \dots \bX_L\}$. For all  $\bs \in [0,\infty)^L$, the mutual information function $I_{\underline{\bX}}(\bs)$ is defined according to
\begin{align*}
I_{\underline{\bX}}(\bs) & = \frac{1}{N}\,  I\big(\underline{\bX} ; \underline{\widetilde{\bX}}(\bs)\big)
\end{align*}
where $\underline{\widetilde{\bX}}(\bs) = \{ \widetilde{\bX}_1(s_1), \dots, \widetilde{\bX}_L(s_L)\}$ are  obtained under independent Gaussian noise. 
In this case, it is possible that the vectors have different dimensions and thus the parameter $N$ should be regarded as a global normalization term. 
The MMSE function $M_{\underline{\bX}}(\bs)$ is defined to be the vector-valued function whose $\ell$-th entry corresponds to the MMSE in the $\ell$-th vector: 
\begin{align*}
\left[ M_{\underline{\bX}}(\bs) \right]_\ell & = \frac{1}{N} \mmse(\bX_\ell \mid \underline{ \widetilde{\bX}}(\bs)).
\end{align*}
By the I-MMSE relationship, the MMSE function is equal to one half  the gradient of the mutual information function: 
\begin{align*}
M_{\underline{\bX}}(\bs) & =  \frac{1}{2} \nabla I_{\underline{\bX} }(\bs),
\end{align*}
The mutual information function $I_{\underline{\bX}}(\bs)$ is concave \cite{payaro:2009}, and the multivariate Legendre function is given by
\begin{align*}
I^*_{\underline{\bX}}(\bu) & = \sup_{s \in [0,\infty)^L} \left( I_{\underline{\bX} \triangle \underline{\bY} \mid \underline{\haar}}(\bs)  - \frac{ 1}{2} \langle \bu, \bs \rangle  \right).
\end{align*}

\subsection{The integrated R-transform and its dual}

The mutual information and MMSE functions corresponding to multivariate Gaussian distributions can be expressed in terms of the Stieltjes  transform and the R-transform from random matrix theory~\cite{voiculescu:1991, tulino:2004}. The Stieltjes  transform of the random variable $X$ is defined by
\begin{align}
\ST_{X}(t) & = \ex{\frac{1}{ X - t}}.  \label{eq:Stieltjes}
\end{align}
This expectation is well-defined for all $t$ outside the support of $X$. For the purposes of this paper, we apply this transform to nonnegative random and restrict the domain to the negative reals $t \in (-\infty,0)$. The Stieltjes  transform associated with an $N \times N$ symmetric  random matrix $\bK$ is defined according to 
\begin{align}
\ST_{\bK}(t) = \frac{1}{N}  \ex{ \gtr\left(  \left( \bK - t I \right)^{-1}  \right) }.
\end{align}
Note that this expression is equivalent to \eqref{eq:Stieltjes} when $X$ is distributed according to the empirical spectral distribution of $\bK$, that is $X$ is equal to $\lambda_i(\bK)$ with probability $1/N$. 

To see the connection with the MMSE function defined in the previous sections, observe that if $\bX \sim \normal(0, \bK^{-1})$  where $\bK$ is a positive definite random matrix  then
\begin{align*}
M_{\bX \mid \bK}(s) & = \frac{1}{N}\ex{ \gtr\left( \left( s I + \bK \right)^{-1}    \right)}  = C_{\bK}(-s) .
\end{align*} 

It can can be verified that the  Stieltjes  transform is a bijection from its domain onto its image, and thus the functional inverse $C_{X}^{-1}(\cdot)$ is well-defined.  The R-transform of the  random variable $X$ is defined by
\begin{align}
\FR_{X}(z) & =  \ST^{-1}_{X}(-z) - \frac{1}{z}.
\end{align}

For the purposes of this paper we focus on the integrated R-transform, which is defined according to
\begin{align}
\IR_{X}(t) & = \frac{1}{2} \int_0^t \FR_{X}(-z) \, \dd z.  \label{eq:IR}
\end{align}
It can be verified that $\IR_{X}(t)$ is concave with
\begin{align*}
\IR_{a X}(t) & = \IR_{ X}(a t) , \qquad a > 0.
\end{align*}
The  Legendre transform of the integrated R-transform is defined according to 
 \begin{align}
\IR^*_{X}(u) & = \sup_{t} \left( \IR_{X}(t)  - \frac{1}{2} u \, t  \right). 
\end{align}

\begin{example}[Point-mass distribution] The transforms associated with a point-mass distribution at $\lambda$ are given by
\begin{align*}
\ST(t) & = \frac{1}{\lambda -  t} , 
& \FR(z) 
& = \lambda \\ 
J(t) & = \frac{1}{2}  \lambda t , & 
\IR^*(u) & = \begin{dcases}
0 , & u  = \lambda\\
+ \infty , & u \ne \lambda.
\end{dcases}
\end{align*}
\end{example}

\begin{example}[Bernoulli distribution]Let $\bA$ be an $M \times N$ matrix $(M \le N)$ distributed uniformly over the Stiefel manifold, i.e., the set of all matrices with orthogonal rows, i.e., $\bA \bA^T = I_M$. The empirical spectral distribution $\bA^T \bA$ is Bernoulli with parameter $\beta  = M/K \in (0,1]$, and the associated transforms are given by
\begin{align*}
\ST(t) & = \frac{\beta}{1- t} - \frac{1-\beta}{t} \\
\FR(z) & = \frac{ z-1  +  \sqrt{ (1-z)^2 +  4 \beta z}}{ 2 z} \\
\IR^*(u) & = \frac{\beta}{2} \log\left( \frac{ \beta}{u} \right) + \frac{(1-\beta)}{2} \log\left( \frac{ 1- \beta}{ 1- u } \right).
\end{align*}
\end{example}

\begin{example}[Mar\v{c}enko-Pastur distribution]
Let $\bA$ be an $M \times N$ random matrix whose entries are IID zero-mean random variables with variance $1/ N$.  If  $M, N \to \infty$ with $M/N \to \beta \in (0,\infty)$, then the empirical spectral distribution of $\bA^T \bA$ converges weakly to the Mar\v{c}enko-Pastur distribution. The transforms associated with this distribution are given by
\begin{align*}
\ST(t) &  =  \frac{ -1 + \beta - t - \sqrt{ t^2 - 2 (1+ \beta) t + (1 -\beta)^2}}{2 t}\\
\FR(z) &= \frac{\beta}{1 - z}\\
 \IR(t) & = \frac{\beta}{2} \log(1 + t)\\ 
 \IR^*(u) & = \frac{\beta}{2} \left(\log\left( \frac{ \beta}{u} \right) + \frac{u}{\beta} - 1 \right).
\end{align*}
\end{example}

\subsection{Examples of transforms}

\begin{example}[Gaussian Prior]
If $\bX \sim \normal(0, \bK^{-1})$  where $\bK$ is a positive definite random matrix  then
\begin{align*}
I^*_{\bX \mid \bK}(u)  & =   \frac{1}{2}\log\left( \frac{ \sigma^2}{u} \right)  + J_{\bK}(u) - J_{\bK}(\sigma^2) \\
  \IM_{\bX \mid \bK}(u) &= \frac{1}{u} - \FR_{\bK}(-u),
\end{align*}
with $\sigma^2 = \frac{1}{N}\ex{ \gtr(\bK^{-1})}$.  Furthermore, if $\bK$ is deterministic and proportional to the identity matrix, then 
\begin{align}
 I^*_{\bX}(u) & =  \frac{1}{2} \left(  \log\left( \frac{\sigma^2}{u} \right)  + \frac{u}{\sigma^2} - 1 \right)\label{eq:Istar_G_prior}\\
\IM_{\bX}(u) & = \frac{1}{u} - \frac{1}{\sigma^2}. \label{eq:IM_G_prior} 
\end{align}
\end{example}

\begin{example}[Gaussian Linear Model] Suppose that
\begin{align*}
\bX \sim \normal(0, \sigma^2 I), \qquad \bZ = \bA \bX, \qquad  \bY =\bZ +\bW,
\end{align*}
where $\bA$ is an $M \times N$ random matrix and $\bW \sim \normal(0, I)$ is standard Gaussian noise. Then, 
\begin{align*}
I^*_{\bX, \bZ \mid \bA}(u, v)   & = I_{\bX}^*(u) +  \IR^*_{\bA^T \bA}\left(v/u\right)\\
I^*_{\bX  \triangle \bY \mid \bA}(u) & =  I^*_{\bX}(u)  + \IR_{\bA^T \bA}(u), 
\end{align*}
where $I_{\bX}^*(u)$ is given in \eqref{eq:Istar_G_prior}.
\end{example}

\begin{example}
 If $\bX \sim \normal(0, \sigma^2 I)$ and $\bY = \sqrt{ \lambda} \bX + \bW$ where $\bW \sim \normal(0, I)$, then 
\begin{align*}
& I^*_{\bX, \bY}(u, v) =  I_{\bX}^*(u) \\
 &\quad  +   \frac{1}{2} \left( \log\left(  \frac{1 +  \sqrt{1 + 4 u v \lambda}}{2 v} \right)   -\sqrt{ 1 + 4 u v \lambda}   + \lambda u   +   v   \right).
\end{align*}
where $I_{\bX}^*(u)$ is given in \eqref{eq:Istar_G_prior}.
\end{example}

\section{Networks of GLMs}\label{sec:MLnetworks}

\subsection{Problem Formulation} 
Consider a rooted tree with nodes indexed by the set $V = V_\text{var} \cup V_\text{obs}$ where $V_\text{var}$ and $V_\text{obs}$ are disjoint sets corresponding to vectors of unknown variables  and observations, respectively. To simplify the exposition we will assume that the root node $1$ belongs to the variable set and that all the nodes indexed by the observation set are terminal nodes (leafs) in the tree. For each index $\ell \in V \backslash\{1\}$ let $\pi(\ell)$ denote the parent of node $\ell$. On this tree, a network of GLMs is defined according to 
\begin{alignat}{3}
\bX_1 &\sim P_1(\bx)  \label{eq:orth_network_a} \\
\bZ_\ell & = \bA_\ell \bX_{\pi(\ell)}, & \qquad & \ell \in V\backslash\{1\}\\
\bX_\ell \mid \bZ_\ell &\sim P_\ell(\bx \mid \bz )  & \qquad & \ell \in V_\text{var}\backslash\{1\}\\
\bY_\ell \mid \bZ_\ell &\sim P_\ell(\by  \mid \bz )  & \qquad & \ell \in V_\text{obs}. \label{eq:orth_network_d} 
\end{alignat}
The dimensions of the vectors and matrices are given by
\begin{align*}
\bX_\ell \in \reals^{N_\ell}, \quad \bA_\ell  \in \reals^{M_\ell \times N_{\pi(\ell)}} , \quad \bZ_\ell \in \reals^{M_{\ell}}.
\end{align*}

We use the convention that the mutual information and MMSE functions associated the network are defined with respect to a global normalization term $N$. For example, if a vector in the $\ell$-th stage consists of $N_\ell$ independent copies of a random variable $X$, then the corresponding mutual information functions defined with respect to $N$ are given by
\begin{align*}
I_{\bX_\ell}(s) & = \alpha_\ell \, I_X(s) & M_{\bX_\ell}(s) & = \alpha_\ell \,  M_X(s)\\
I^*_{\bX_\ell}(u) & = \alpha_\ell \, I^*_X\left( u / \alpha_\ell \right) & \IM_{\bX_\ell}(u) & =   \IM_X\left(u / \alpha_\ell \right),
\end{align*}
where $\alpha_\ell = N_\ell/ N$.  The relationships between the mutual information and MMSE functions described in Sections~\ref{sec:transforms}  are unaffected by this choice of normalization.  

Our construction of the potential function consists of the following terms: 
\begin{alignat}{3}
\Psi_1(u) & =  I_{\bX_1}^*(u_1) \\
\Psi_\ell(u,v) & =  I_{\bZ_{\ell}, \bX_\ell}^*(u, v)  - I_{\bZ_\ell}^*(u) & \qquad & \ell \in V_\text{var}\backslash\{1\}\\
\Phi_\ell(u) & =  I_{\bZ_\ell \triangle \bY_\ell}^*(u) -  I_{\bZ_\ell}^*(u)& \qquad & \ell \in V_\text{obs}.
\end{alignat}
The term $\Psi_1(u)$ depends on the prior distribution associated with the root node. The term $\Psi_\ell(u,v)$ depends on the joint distribution of  the input-output pair $(\bZ_\ell, \bX_\ell)$ associated with an unobserved node. The term $\Phi_\ell(u)$ depends on the joint distribution of the input-output pair $(\bZ_\ell, \bY_\ell)$ associated with an  observed node. 

Under the assumption that $\bA_\ell$ is orthogonally invariant, the distribution of $\bZ_\ell$ is also orthogonally invariant, and thus
\begin{align*}
\bZ_\ell \overset{\text{dist}}{=}\left \|\bX_{\pi(\ell)} \right\| \, \sqrt{ \left( \bA_\ell^T \bA_\ell \right)_{1,1} } 
\, \bV_\ell, 
\end{align*}
where $\bV_\ell$ is distributed uniformly on the sphere of radius one. Consequently, if $\frac{1}{N} \|\bX_{\pi(\ell)}\|^2$ and $\left( \bA_\ell^T \bA_\ell \right)_{1,1}$  are close to their expectations with high probability, then the distribution of $\bZ_\ell$ is approximately isotropic Gaussian $\normal(0, \tau^2_\ell I)$ with 
\begin{align*}
\tau^2_\ell& =\ex{  \frac{1}{N}\left \|\bX_{\pi(\ell)}\right\|^2} \ex{ \frac{1}{N} \left\| \bA_\ell \right \|_F^2}.
\end{align*}

\begin{definition} The potential function associated with the tree network of GLMs in  \eqref{eq:orth_network_a}--\eqref{eq:orth_network_d} is defined according to: 
\begin{align*}
\cF_N(\bu,\bv) & =  \Psi_1(u_1) + \sum_{\ell \in V_\text{var} \backslash\{1\}} \Psi_{\ell}(v_{\pi(\ell)}, u_\ell) \notag \\
&  + \sum_{\ell \in V_\text{obs}} \!  \Phi_{\ell}(v_{\pi(\ell)}) + \sum_{\ell \in V \backslash\{1\}} \! \! \! \IR^*_{\bA_\ell^T \bA_\ell}\left( \frac{v_\ell}{ u_{\pi(\ell)}} \right).
\end{align*}
The number of entries in $\bu = \{ u_\ell\} $ is equal to the number of unobserved nodes and the number of entries in $\bv = \{ v_\ell\} $ is equal to the number of GLMs.  The domain is given by the intersection of the domains of the individual terms. 
\end{definition}

We also define a compact version of the potential function $\widetilde{\cF}_N(\bu)$ corresponding to the minimum of $\cF_N(\bu, \bv)$ with respect to $\bv$. This function can be expressed as
\begin{align*}
\widetilde{\cF}_N(\bu) 
& = \Psi(u_1) + \sum_{\ell \in V_\text{var} \backslash\{1\}} \widetilde{\Psi}_{\ell}(u_{\pi(\ell)}, u_\ell) + \sum_{\ell \in V_\text{obs}} \widetilde{\Phi}_{\ell}(u_{\pi(\ell)}) 
\end{align*}
where
\begin{align*}
\widetilde{\Psi}_{\ell}(u, u') & = \min_{v} \left(  \IR^*_{\bA_\ell^T \bA_\ell}\left(v/u \right) + \Psi_\ell(v, u')  \right)\\
\widetilde{\Phi}_{\ell}(u) & = \min_{v} \left(  \IR^*_{\bA_\ell^T \bA_\ell}\left(v/u \right) + \Phi_\ell(v)  \right).
\end{align*}
Note that in the special case where $\bA_\ell$ is an orthogonal matrix, we have $\widetilde{\Psi}_{\ell}(u, u') = \Psi_{\ell}(u, u') $ and $\widetilde{\Phi}_{\ell}(u) = \Phi_\ell(u)$.

\subsection{Converging sequences} 

We focus on sequences of problems indexed by $N$ in which the network is fixed, while the dimensions $\{M_\ell, N_\ell\}$ increase to infinity with
\begin{align*}
\lim_{N\to \infty} N_\ell /  N  =  \alpha_\ell  , \qquad \lim_{N\to \infty} M_\ell /  N  =  \beta_\ell  
\end{align*}
where  $\alpha_\ell, \beta_\ell \in (0,\infty).$

\begin{definition} A sequence of problems is said to be \textit{converging} if there exists an absolutely  continuous  function $\cF(\bu, \bv)$ such that
\begin{align*}
\lim_{N \to \infty} \cF_N(\bu, \bv) =  \cF(\bu, \bv) ,
\end{align*}
for all $(\bu, \bv)$ in the domain of $\cF(\bu,\bv)$. 
\end{definition}

\begin{definition} The potential function formalism associated with a converging sequence of tree network of GLMs is said to be \textit{asymptotically exact} if the following conditions holds: 
\begin{enumerate}[(i)]
\item  The mutual information satisfies 
\begin{align*}
\lim_{N \to \infty} \frac{1}{N} I(\underline{\bX} ; \underline{\bY} \mid  \underline{\bA})  & = \min_{\bu, \bv} \cF(\bu, \bv) .
\end{align*}
\item If $\cF(\bu, \bv)$ has a unique global minimum at $(\bu^*, \bv^*)$, then the MMSE satisfies
\begin{align*}
\lim_{N \to \infty} \frac{1}{N} \mmse(\bX_\ell \mid \underline{\bY}, \underline{\bA}) & = u^*_\ell, \qquad \ell \in V_\mathrm{obs}
\\
\lim_{N \to \infty} \frac{1}{N} \mmse(\bZ_\ell \mid \underline{\bY}, \underline{\bA}) & = v^*_\ell, \qquad \ell \in V\backslash \{1\}.
\end{align*}
\end{enumerate}
\end{definition}

\subsection{Results} 

\begin{assumption}[Orthogonally Invariant Matrices]\label{assumption:invariant_matrices} The matrices $\{\bA_\ell\}$ are drawn independently from orthogonally invariant distributions.  Furthermore, as $N$ increases to infinity, $\ex{ \lambda^2_\text{max}(\bA_\ell^T \bA_\ell)}$ is bounded uniformly and the empirical spectral distribution of $\bA_\ell^T \bA_\ell$ converges weakly and almost surely to a compactly supported probability measure $\nu_\ell$.  
\end{assumption}

\begin{assumption}[Gaussian Case]\label{assumption:Gaussian_case} The prior distribution is Gaussian $\bX_1 \sim   \normal(0,\sigma_1^2 I)$ and every channel in the network corresponds to additive Gaussian noise $P_{\ell}( \cdot \mid \bz) = \normal( \bz,  I)$. 
\end{assumption}

\begin{theorem}
Under Assumptions \ref{assumption:invariant_matrices} and \ref{assumption:Gaussian_case}, the sequence of problems is converging and the potential function formalism is asymptotically exact. Furthermore, the asymptotic potential function $\cF(\bu, \bv)$ is convex.
\end{theorem}

\begin{assumption}[Separable Case]\label{assumption:Seperable_case} $\quad$ 
\begin{enumerate}[(i)]
\item The entries of $\bX_1$ are IID copies of a random variable $X$ with finite fourth moment. 
\item The channels are separable and have a Gaussian component, i.e., 
\begin{align}
 P_{\ell}(\bx \mid \bz) & = \prod_{i=1}^{N_\ell} \int \phi_{\eps_\ell}\left( u\right) \,  Q_\ell\left(x_i - u  \mid z_i\right)  \, \dd u ,
\end{align}
where  $Q_\ell(\cdot \mid z)$ is a fixed probability measure on $\reals$ with uniformly bounded fourth moment,  $\phi_{\sigma^2}(u)$ is the density of a Gaussian $\normal(0, \sigma^2)$ distribution, and the variance terms $\{\eps_\ell\}$ are strictly positive. 
\end{enumerate}
\end{assumption} 

\begin{theorem}
Under Assumptions \ref{assumption:invariant_matrices} and \ref{assumption:Seperable_case}, the sequence of problems is converging. Furthermore:
\begin{enumerate}[(i)]
\item In the case of a sequential network, the stationary points of the potential function $\cF(\bu, \bv)$ characterize the fixed-points of the state evolution of the ML-VAMP algorithm introduced by Fletcher and Rangan~\cite{fletcher:2017a}.

\item In the case of a serial network with IID Gaussian matrices, the potential function formalism recovers the formulas for the mutual information and MMSE obtained by Manoel et al.~\cite{manoel:2017a} using the replica method. 

\item In the case of the GLM with an IID Gaussian matrix, the  potential function formalism is asymptotically exact. In particular, the formulas for the mutual information and MMSE  match the expressions given by Barbier et al.~\cite{barbier:2017c}.

\item In the case of a standard linear model,   the potential function formalism recovers the formulas for the mutual information and MMSE obtained by Tulino et al.~\cite{tulino:2013} using the replica method. 
\end{enumerate}

\end{theorem}

\section{Conclusion}
This paper describes a new method for analyzingt the statistical properties of multilayer networks. The main assumption made throughout is that the matrices are drawn independently from orthogonally invariant distributions. Our method is rigorous for a variety of problems, including the special case of Gaussian networks. More generally, our method provides precise conjectures for non-Gaussian networks. An important direction for future work is to study the necessary and sufficient conditions for the potential function formalism to be asymptotically exact.

\bibliographystyle{IEEEtran}
\bibliography{multilayer_glm_arxiv.bbl} 

\begin{thebibliography}{10}
\providecommand{\url}[1]{#1}
\csname url@samestyle\endcsname
\providecommand{\newblock}{\relax}
\providecommand{\bibinfo}[2]{#2}
\providecommand{\BIBentrySTDinterwordspacing}{\spaceskip=0pt\relax}
\providecommand{\BIBentryALTinterwordstretchfactor}{4}
\providecommand{\BIBentryALTinterwordspacing}{\spaceskip=\fontdimen2\font plus
\BIBentryALTinterwordstretchfactor\fontdimen3\font minus
  \fontdimen4\font\relax}
\providecommand{\BIBforeignlanguage}[2]{{%
\expandafter\ifx\csname l@#1\endcsname\relax
\typeout{** WARNING: IEEEtran.bst: No hyphenation pattern has been}%
\typeout{** loaded for the language `#1'. Using the pattern for}%
\typeout{** the default language instead.}%
\else
\language=\csname l@#1\endcsname
\fi
#2}}
\providecommand{\BIBdecl}{\relax}
\BIBdecl

\bibitem{voiculescu:1991}
D.~Voiculescu, ``Limit laws for random matrices and free products,''
  \emph{Inventiones mathematicae}, vol. 104, no.~1, pp. 201--220, 1991.

\bibitem{tulino:2004}
A.~Tulino and S.~Verd\'{u}, \emph{Random Matrix Theory and Wireless
  Communications}.\hskip 1em plus 0.5em minus 0.4em\relax Hanover, MA: now
  Publisher Inc., 2004.

\bibitem{mezard:2009}
M.~M\'{e}zard and A.~Montanari, \emph{Information, physics, and
  computation}.\hskip 1em plus 0.5em minus 0.4em\relax Oxford University Press,
  2009.

\bibitem{tanaka:2002}
T.~Tanaka, ``A statistical-mechanics approach to large-system analysis of
  {CDMA} multiuser detectors,'' \emph{IEEE Trans.\ Inform.\ Theory}, vol.~48,
  no.~11, pp. 2888--2910, Nov. 2002.

\bibitem{guo:2005}
D.~Guo and S.~Verd\'{u}, ``Randomly spread {CDMA}: Asymptotics via statistical
  physics,'' \emph{IEEE Trans.\ Inform.\ Theory}, vol.~51, no.~6, pp.
  1983--2010, Jun. 2005.

\bibitem{guo:2009}
D.~Guo, D.~Baron, and S.~Shamai, ``A single-letter characterization of optimal
  noisy compressed sensing,'' in \emph{Proc.\ Annual Allerton Conf.\ on
  Commun., Control, and Comp.}, Monticello, IL, Oct. 2009.

\bibitem{reeves:2012}
G.~Reeves and M.~Gastpar, ``The sampling rate-distortion tradeoff for sparsity
  pattern recovery in compressed sensing,'' \emph{IEEE Trans.\ Inform.\
  Theory}, vol.~58, no.~5, pp. 3065--3092, May 2012.

\bibitem{tulino:2013}
A.~Tulino, G.~Caire, S.~Verd\'{u}, and S.~Shamai, ``Support recovery with
  sparsely sampled free random matrices,'' \emph{IEEE Trans.\ Inform.\ Theory},
  vol.~59, no.~7, pp. 4243--4271, Jul. 2013.

\bibitem{manoel:2017a}
A.~Manoel, F.~Krzakala, M.~M\'ezard, and L.~Zdeborov\'a, ``Multi-layer
  generalized linear estimation,'' in \emph{Proc.\ IEEE Int.\ Symp.\ Inform.\
  Theory}, Aachen, Germany, 2017, pp. 2098--2102.

\bibitem{korada:2010}
S.~B. Korada and N.~Macris, ``Tight bounds on the capicty of binary input
  random {CDMA} systems,'' \emph{IEEE Trans.\ Inform.\ Theory}, vol.~56,
  no.~11, pp. 5590--5613, Nov. 2010.

\bibitem{huleihel:2017}
W.~Huleihel and N.~Merhav, ``Asymptotic {MMSE} analysis under sparse
  representation modeling,'' \emph{Signal Processing}, vol. 131, pp. 320--332,
  2017.

\bibitem{reeves:2016a}
G.~Reeves and H.~D. Pfister, ``The replica-symmetric prediction for compressed
  sensing with {G}aussian matrices is exact,'' in \emph{Proc.\ IEEE Int.\
  Symp.\ Inform.\ Theory}, Barcelona, Spain, Jul. 2016, pp. 665 -- 669.

\bibitem{barbier:2016}
J.~Barbier, M.~Dia, N.~Macris, and F.~Krzakala, ``The mutual information in
  random linear estimation,'' in \emph{Proc.\ Annual Allerton Conf.\ on
  Commun., Control, and Comp.}, Monticello, IL, 2016.

\bibitem{barbier:2017c}
J.~Barbier, F.~Krzakala, N.~Macris, L.~Miloane, and L.~Zdeborov\'a, ``Phase
  transitions, optimal errors and optimality of message-passing in generalized
  linear models,'' Aug. 2017, [Online]. {A}vailable
  \url{https://arxiv.org/abs/1708.03395}.

\bibitem{donoho:2009a}
D.~L. Donoho, A.~Maleki, and A.~Montanari, ``Message-passing algorithms for
  compressed sensing,'' \emph{Proceedings of the National Academy of Sciences},
  vol. 106, no.~45, pp. 18\,914--18\,919, Nov. 2009.

\bibitem{bayati:2011}
M.~Bayati and A.~Montanari, ``The dynamics of message passing on dense graphs,
  with applications to compressed sensing,'' \emph{IEEE Trans.\ Inform.\
  Theory}, vol.~57, no.~2, pp. 764--785, Feb. 2011.

\bibitem{rangan:2011}
S.~Rangan, ``Generalized approximate message passign for estimation with random
  linear mixing,'' in \emph{Proc.\ IEEE Int.\ Symp.\ Inform.\ Theory}, St.
  Petersburg, Russia, 2011, pp. 2174--2178.

\bibitem{cakmak:2014a}
B.~\c{C}akmak, O.~Winther, and B.~H. Fleury, ``{S}-{AMP}: {A}pproximate message
  passing for general matrix ensembles,'' in \emph{Proc.\ IEEE Inform.\ Theory
  Workshop}, Hobart, TAS, Australia, 2014, pp. 192--196.

\bibitem{rangan:2016a}
S.~Rangan, P.~Schniter, and A.~K. Fletcher, ``Vector approximate message
  passing,'' Oct. 2016, [Online]. {A}vailable
  \url{https://arxiv.org/abs/1610.03082}.

\bibitem{schniter:2016b}
P.~Schniter, S.~Rangan, and A.~K. Fletcher, ``Vector approximate message
  passing for the generalized linear model,'' Dec. 2016, [Online]. {A}vailable
  \url{https://arxiv.org/abs/1612.01186}.

\bibitem{fletcher:2017a}
A.~K. Fletcher and S.~Rangan, ``Inference in deep networks in high
  dimensions,'' 2017, [Online]. {A}vailable
  \url{https://arxiv.org/abs/1706.06549}.

\bibitem{opper:2005}
M.~Opper and O.~Winther, ``Expectation consistent approximate inference,''
  \emph{Journal of Machine Learning Research}, vol.~6, pp. 2177--2204, 2005.

\bibitem{reeves:2017c}
G.~Reeves, ``Conditional central limit theorems for {G}aussian projections,''
  in \emph{Proc.\ IEEE Int.\ Symp.\ Inform.\ Theory}, Aachen, Germany, Jun.
  2017, pp. 3055--3059.

\bibitem{tse:1999}
D.~N.~C. Tse and S.~Hanly, ``Linear multiuser receivers: Effective
  interference, effective bandwith and user capacity,'' \emph{IEEE Trans.\
  Inform.\ Theory}, vol.~45, pp. 641--657, Mar. 1999.

\bibitem{verdu:1999}
S.~Verd\'{u} and S.~Shamai, ``Spectral efficiency of cdma with random
  spreading,'' \emph{IEEE Trans.\ Inform.\ Theory}, vol.~45, pp. 622--640, Mar.
  1999.

\bibitem{guo:2011}
D.~Guo, Y.~Wu, S.~Shamai, and S.~Verd\'{u}, ``Estimation in {G}aussian noise:
  {P}roperties of the minimum mean-square error,'' \emph{IEEE Trans.\ Inform.\
  Theory}, vol.~57, no.~4, pp. 2371--2385, Apr. 2011.

\bibitem{guo:2005a}
D.~Guo, S.~Shamai, and S.~Verd\'{u}, ``Mutual information and minimum
  mean-square error in {G}aussian channels,'' \emph{IEEE Trans.\ Inform.\
  Theory}, vol.~51, no.~4, pp. 1261--1282, Apr. 2005.

\bibitem{verdu:2010}
S.~Verd\'{u}, ``Mismatched estimation and relative entropy,'' \emph{IEEE
  Trans.\ Inform.\ Theory}, vol.~56, no.~8, pp. 3712 -- 3720, Aug. 2010.

\bibitem{payaro:2009}
M.~Payar\'{o} and D.~Palomar, ``Hessian and concavity of mutual information,
  differential entropy, and entropy power in linear vector {G}aussian
  channels,'' \emph{IEEE Trans.\ Inform.\ Theory}, vol.~55, no.~8, pp.
  3613--3628, 2009.

\end{thebibliography}

\end{document}